\documentclass[a4paper,aps,english,twocolumn,floatfix,amsfonts,amssymb,superscriptaddress,longbibliography]{revtex4-1}
\usepackage{graphicx}
\usepackage{epstopdf} 
\usepackage{dcolumn}
\usepackage{bm}
\usepackage{bbm,bbold}
\usepackage{color}
\usepackage{xcolor, soul}
\sethlcolor{green}
\usepackage{amsfonts}
\usepackage{amsmath}
\usepackage{mdframed}
\usepackage[normalem]{ulem}
\usepackage{mathrsfs}   
\usepackage[none]{hyphenat}
\usepackage[percent]{overpic}
\usepackage[colorlinks=true,citecolor=blue]{hyperref}
\hypersetup{colorlinks=true,citecolor=blue,linkcolor=blue,urlcolor=blue}

\begin{document}
%\title{Pseudo-Gauge Field Driven Acoustoelectric Current in 2D hexagonal Dirac Materials}
\title{Pseudogauge field driven acoustoelectric current in two-dimensional hexagonal Dirac materials}

\author{Pankaj Bhalla}
	\affiliation{Nordita, KTH Royal Institute of Technology and Stockholm University, Hannes Alfvéns väg 12, 10691 Stockholm, Sweden}
\author{Giovanni Vignale}
	\affiliation{Department of Physics and Astronomy, University of Missouri, Columbia, Missouri 65211, USA}
		\author{Habib Rostami}
	\affiliation{Nordita, KTH Royal Institute of Technology and Stockholm University, Hannes Alfvéns väg 12, 10691 Stockholm, Sweden}
\date{\today}
\begin{abstract}
Using a diagrammatic scheme, we study the acoustoelectric effects in two-dimensional (2D) hexagonal Dirac materials due to the sound-induced pseudo-gauge field. We analyze both uniform and {\em spatially dispersive} currents in response to copropagating and counterpropagating sound waves, respectively. In addition to the longitudinal acoustoelectric current, we obtain an exotic {\em transverse} charge current flowing perpendicular to the sound propagation direction owing to the interplay of transverse and longitudinal gauge field components $j_T\propto A_L A^\ast_T$.
In contrast to the almost isotropic directional profile of the longitudinal uniform current, a highly anisotropic transverse component $j_T\sim\sin(6\theta)$ is achieved that stems from the inherited three-fold symmetry of the hexagonal lattice. However, both longitudinal and transverse parts of the dispersive current are predicted to be strongly anisotropic $\sim\sin^2(3\theta)$ or $\cos^2(3\theta)$.
We quantitatively estimate the pseudogauge field contribution to the acoustoelectric current that can be probed in future experiments in graphene and other 2D hexagonal Dirac materials.
\end{abstract}
\maketitle

\section{Introduction}
The passage of a sound wave through  an electronic system  creates an oscillating electric field which accelerates the charge carriers and generates an electric current. The acoustoelectric effect (AE) is the dc current that arises to second-order in the sound-induced electric field.  This intriguing nonlinear phenomenon was first predicted by Parmenter \cite{parmenter_PR1953} and later discussed by Weinreich \cite{weinreich_1957}. The effect has been observed in different classes of materials such as semiconductors, quantum wires, two-dimensional (2D) electron gas, and heterostructures  \cite{Weinreich_White_1957,wataru_JPSJ1957, weinreich_PR1959, smith_Nat1959, eckstein_JAP1964, willett_PRL1990,Ilisavskii_prl_2001,Gloos_prb_2004,Kreft_prb_2016}. More recently, it has been recognized that  the coupling between the surface acoustic wave (SAW) and electrons in 2D Dirac materials provides an exciting opportunity to investigate charge transport driven by the strain fields associated with propagating SAW \cite{falko_PRB1993, simon_PRB1996, miseikis_APL2012, bandhu_APL2013, bandhu_APL2014, zheng_APL2016, bandhu_NR2016,Poole2017}. In particular, the AE effect of single-layer graphene has been investigated experimentally,
and the AE current has been shown to be tunable by the application of a gate voltage \cite{bandhu_NR2016}.

Traditionally, the  magnitude of the sound-induced uniform direct current  is  obtained from the Weinreich's relation \cite{Weinreich_White_1957, ingebrigtsen_JAP1970,boardman_1990,Xmisc} 
\begin{equation}\label{eq:jAE}
j^{\rm AE}= -\frac{\mu  \Gamma_s I_s }{v_s},
 \end{equation}
where $\Gamma_s$ is the sound attenuation, $I_s$ is the sound intensity, $v_s$ is the sound velocity and $\mu$ is the mobility of the carriers. The mobility has opposite sign in electron and hole doped systems. The sound attenuation in conventional (piezoelectric) semiconductors is  estimated as $\Gamma_s = K^2_p \Omega  (\sigma/\sigma_m)/[1+(\sigma/\sigma_m)^2]$, where $K^2_p$ is related to the piezoelectric constant, $\Omega$ is the sound frequency, and $\sigma$ is the longitudinal conductivity of the system with $\sigma_m$ being a characteristic conductivity constant \cite{wixforth_PRL1986}. 

\begin{figure}[t]
\centering
\includegraphics[width=9cm, height=5cm]{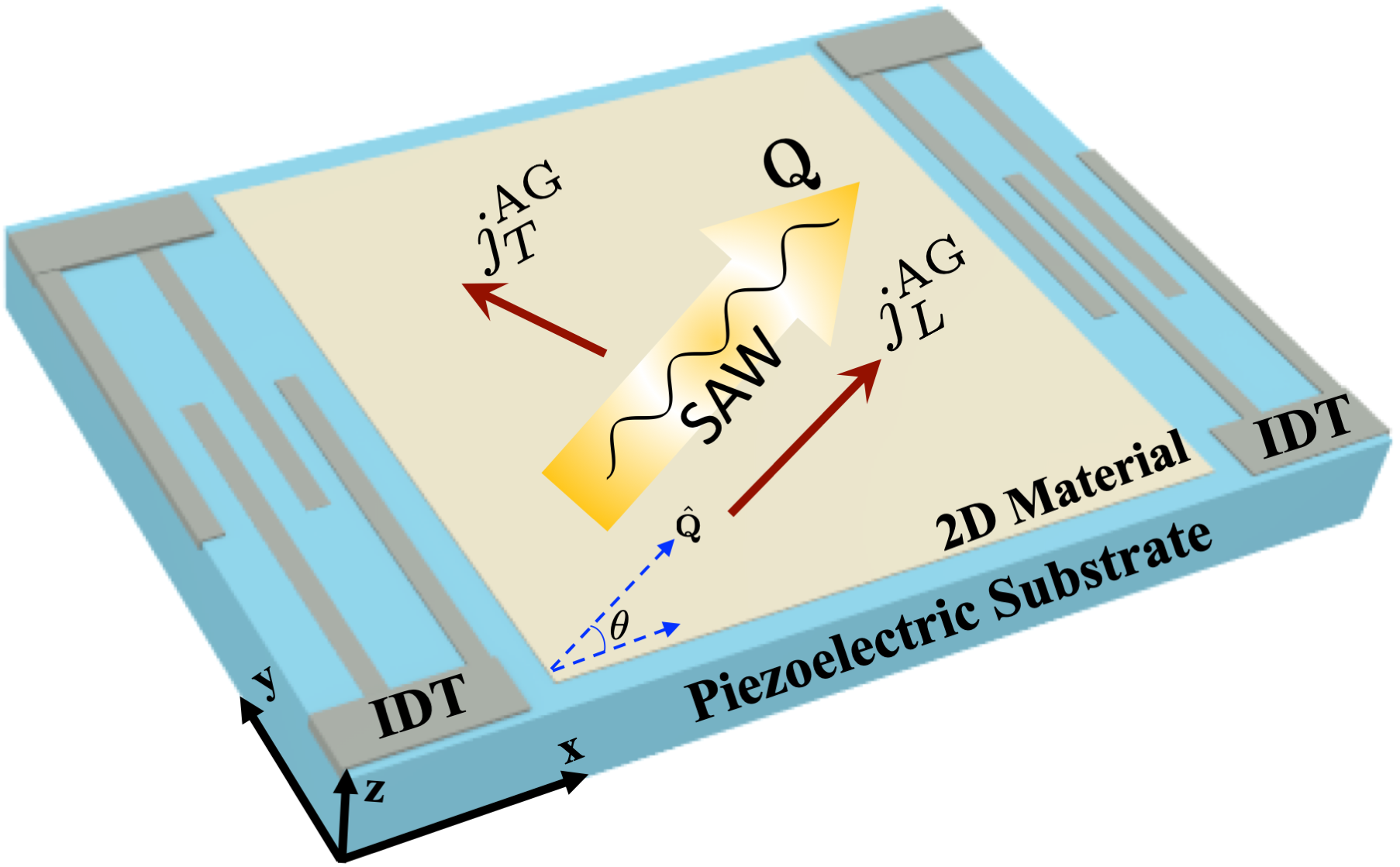}
\caption{Schematic experimental setup demonstrating the generation of the acoustogalvanic (AG) currents due to the second order response to the pseudogauge potential induced by surface acoustic wave (SAW) propagating in the 2D hexagonal Dirac material placed on a piezoelectric substrate. Interdigital transducers (IDTs) convert electric signals into SAW with frequency $\Omega$ propagating along ${\bm Q}$ with an elliptic polarization of the displacement fields, ${\bm u}=u_L \hat {\bm Q} - iu_z \hat{\bm z}$. Here, $j_T^{{\rm AG}}$ and $j_L^{{\rm AG}}$ refer to the longitudinal and transverse AG currents respectively and $\theta$ is an angle obtained by the phonon wave vector with the $x$-direction (zigzag orientation on the hexagonal lattice).  
}
\label{fig:schematic}
\end{figure}

A longitudinal sound wave, with a displacement amplitude $u_L$ and wave vector $Q$, can induce a scalar potential $V = (\Lambda_P-i\Lambda_D Q) u_L$ \cite{mahan_book}, where $\Lambda_D$ and $\Lambda_P$ stand for two distinct contributions namely the deformation and piezoelectric couplings, respectively. Notice that the deformation potential contribution is relatively less relevant in the long wavelength limit. The SAW on the piezoelectric substrate generates a direct AE current in graphene \cite{Hernandez_2018} and is predicted to induce valley acoustoelectric current in transition metal dichalcogenides (TMDs) \cite{kalameitsev_PRL2019}. 
However, in Dirac materials there is a third contribution to the AE current that can be formally modeled as the acoustic analogous of the photogalvanic effect in response to the sound-induced vector potential ${\bm A}$ \cite{Kane_Mele_1997,Suzuura_2002,Sasaki_2005,Katsnelson_ssc_2007,Guinea_np_2010,Vozmediano_2010,Rostami_2013,Rostami_prb_2015,rostami_npj2dm_2018} (pseudogauge phonons).  We refer to this additional contribution as the   {\em acoustogalvanic effect}~\cite{sukhachov_PRL2020}. Similarly, in 3D Dirac materials a sound-induced orbital magnetization is predicted to arise in the second-order response to a sound-induced vector potential. \cite{long_prl_2021}.

Despite many studies on the piezoelectric mechanism of the AE effect in the 2D electron gas and in graphene \cite{falko_PRB1993, simon_PRB1996, miseikis_APL2012, bandhu_APL2013, bandhu_APL2014, zheng_APL2016, bandhu_NR2016,Poole2017} the  analysis of the acoustoelectric effect in 2D hexagonal Dirac materials is not yet complete. In particular, the relevance of gauge phonon for the AE effect in graphene has not been discussed to the best of our knowledge. Our aim in this paper is to fill this gap by utilizing the diagrammatic second order response method. We discuss the AE effect originating from both scalar and vector potentials. The scalar potential is dynamically screened, while the pseudogauge potential is not screened because it does not generate a charge current in the linear response. 

The conventional AE current, i.e., the current generated by the piezoelectric and deformation potentials flows parallel to the direction of propagation of the SAW, i.e., it is purely longitudinal: $j^{\rm AE}_L || {\bm Q}$. However, the acoustogalvanic (AG) current  contains both longitudinal ($j^{\rm AG}_L$) and transverse ($j^{\rm AG}_T$) components which are parallel and perpendicular to the sound wave vector, respectively -- see the schematic view in Fig.~\ref{fig:schematic}. We obtain a non-trivial dependence of the AG current on the sound propagation direction that stems from the implicit three-fold symmetry of 2D hexagonal Dirac material crystal. We quantitatively analyze the frequency, Fermi energy and angular dependence of the AG current components. Furthermore, in addition to the spatially uniform AG current $j^{\rm AG}_{L,T}$, we study the spatially dispersive current~\cite{ji_NM_2019,ji_Sc2020} $j^{\rm s-AG}_{L,T}$, which exhibits a distinctive dependence on the frequency, the Fermi energy and the propagation angle. The calculated  AG current is in good agreement with  experimental measurements in graphene \cite{bandhu_NR2016}. 
Our calculations are done with the Green's function method, which facilitates the further investigation of the impact of  many-body interactions on the AE effect by systematic many-body perturbation theory and {\em ab initio} approaches. 

\begin{figure*}[t]
    \centering
    \includegraphics[width=18cm]{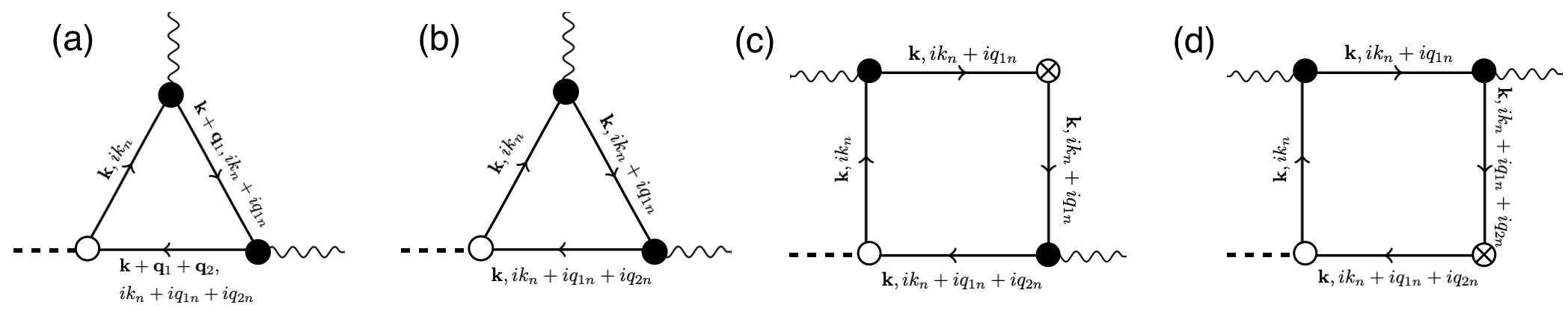}
    \caption{Feynman diagrams for the acoustogalvanic response function in 2D hexagonal Dirac materials. Panel (a) stands for the full nonlinear response function $\chi^{(2)}_{abc}$ with finite phonon wave vector $\bm q$. Panels (b), (c) and (d) are the corresponding diagrams for $X_{abc}$, $Y_{abdc}$ and $Z_{abcd}$ susceptibilities, respectively. Solid circles indicate the electron-phonon coupling vertex, empty circle stands for the coupling of electrons to the electromagnetic (EM) fields (photons). Circle with cross is for a current vertex with vanishing momentum and energy transfer to fermions.  Wave lines are external phonons, dashed line is the output EM wave, and solid lines are the fermionic propagators with corresponding wave vector and Matsubara energies.}
    \label{fig2}
\end{figure*}

\section{Method}\label{sec:method}
A Rayleigh SAW \cite{biryukov1995surface} propagating in the $xy$ plane is an elliptically polarized wave in the $xz$ plane: 
\begin{equation}\label{eq:u}
{\bm u}({\bm r},t) = \frac{1}{2}(u_L \hat{\bm Q} - i u_z \hat{\bm z}) e^{i({\bm Q} \cdot {\bm r} - \Omega t)} + c.c.~, 
\end{equation}
which is propagating on the surface of a piezoelectric substrate along with the phonon wave vector ${\bm Q} = Q(\cos\theta, \sin\theta)$ with $\theta$ being an azimuthal angle. Here, $\theta=0$ corresponds to the $\hat {x}$ direction with a zigzag orientation on the hexagonal lattice in our convention and the SAW dispersion follows the relation $\Omega = v_s Q$. Note that $u_L$ and $u_z$ stand for the longitudinal and normal displacement amplitudes, respectively. Here, we systematically study the direct charge current induced by the sound wave. 

We consider the total Hamiltonian of the system which encapsulates three main mechanisms that contribute to the AE current originating from different sources of Dirac electron's coupling to acoustic phonons
\begin{align}
{\cal H} = v_F \hat {\bm \sigma} \cdot ({\bm p}+ e{\bm A}^{(\tau)}({\bm r},t))+ V({\bm r},t).
\end{align}
Here $v_F$ is the Fermi velocity, $p$ is the momentum of an electron, $\sigma_i$ refers to the Pauli matrices, and $V= \Lambda_D (u_{xx}+u_{yy}) + \Lambda_P u_L$ is a scalar deformation potential that describes the coupling of acoustic phonons to electrons in 2D hexagonal Dirac materials, such as graphene, where $u_{ij}=(\partial_i u_j+\partial_j u_i + \partial_i h \partial_j h)/2$ stands for the strain tensor components in terms of the displacement vector ${\bm u} = (u_x,u_y,h)$ having $h \equiv u_z$ as the normal component of the displacement. The substrate induced piezoelectric potential scales with phonon displacement field $u_{L}$ while the deformation potential scales with $i Q u_{L}$. Therefore, at low frequency the more relevant mechanism is the inter-layer coupling of electrons of graphene to the piezoelectric induced polarization in the substrate.
  
Acoustic deformation also generates a pseudogauge field ${\bm A}^{(\tau)}$ with $\tau=\pm$ indicating two time-reversal counterparts at two valleys with opposite chiralities. In order to comply with the overall time-reversal symmetry, such a gauge field requires opposite sign for two chiralities: 
${\bm A}^{(\tau)} =\tau {\bm A}=\tau A_0 (u_{xx}-u_{yy},-2u_{xy})$ \, \cite{Kane_Mele_1997,Suzuura_2002,Sasaki_2005,Katsnelson_ssc_2007,Guinea_np_2010,Vozmediano_2010,Rostami_2013,Rostami_prb_2015},
where $A_0={\hbar\beta}/{e a}$ in which $a$ is the lattice constant, and the dimensionless parameter $\beta\sim 1$ is the Gruneisen's parameter indicating the strength of the electron-phonon coupling.
Using Eq.~(\ref{eq:u}) and neglecting second order terms in $u_z$, the sound-induced dynamical vector potential reads 
\begin{align}
{\bm A}({\bm r},t) =\frac{1}{2} (A_{L}  \hat {\bf Q} + A_{T} \hat{\bm \theta} ) e^{i({\bm Q} \cdot {\bm r} - \Omega t)}  + c.c.\, ,
\end{align}
where the vector potential is given in terms of longitudinal and transverse components $(A_{L} ,A_{T} )= iA_0(Q u_{L}) (\cos(3\theta),-\sin(3\theta))$.
Note that $\hat{\bm \theta}=\partial_\theta \hat{\bm Q}$ is the azimuthal unit vector transverse to $\hat{\bm Q}$ in the momentum space. Unlike the displacement field of the surface acoustic phonon, the sound-induced gauge field is not an elliptically polarized wave. The vector field stemming from the out-of-plane displacement is proportional to the $u_z^2$ due to $\partial_x h\partial_y h$ and therefore its impact on the AG current is less relevant. 

The time and space dependent strain induces scalar potential and pseudogauge field leading to the effective electromagnetic fields ${\bm E}= -\partial_{\bm r}V-\partial_t {\bm A}$ and ${\bm B} = \partial_{\bm r} \times {\bm A}$.
In the second order response to sound-induced fields, the acoustoelectric current or the rectification current formally follows  
\begin{align}
&{\cal J}^{(2)}_\lambda({\bm r},t) =\sum_\tau \sum_{{\bm q}_1,{\bm q}_2}
\sum_{\omega_1,\omega_2}
 \sum_{\mu,\nu}  
\chi^{(2)}_{\lambda\mu\nu}({\bm q}_1,\omega_1,{\bm q}_2,\omega_2)
\nonumber\\& 
 {\cal A}_\mu({\bm q}_1,\omega_1)  {\cal A}_\nu({\bm q}_2,\omega_2)  e^{-i(\omega_1+\omega_2)t}  e^{i ({\bm q}_1+{\bm q}_2)\cdot {\bm r}},
\end{align}
where $\chi^{(2)}_{\lambda\mu\nu}$ is the nonlinear acoustoelectric response function. 
Note that ${\cal J}^{(2)}=(n^{(2)},{\bm j}^{(2)})$ is the four-vector of nonlinear density $n^{(2)}$ and current ${\bf j}^{(2)}$, ${\cal A}= (V^{\rm sc},\tau{\bm A}^{\rm sc})$ is an effective four-vector potential, where $V^{\rm sc}$ and ${\bm A}^{\rm sc}$ are the self-consistent sound-induced scalar and vector potentials, respectively. The second order current generated by the interplay of scalar and vector potentials, $\tau V^{\rm sc}{\bm A}^{\rm sc}$, cancels out due to opposite contributions of two valleys $\tau = \pm$.  
\subsection{Diagrammatic theory of nonlinear response function}
We consider a 2D Dirac material whose low-energy Hamiltonian reads ${\cal H}({\bf k}) = \hbar v_F (\tau k_x \sigma_x + k_y \sigma_y)$, where $\tau = \pm$ for the valley index $(K, K')$, $v_F$ is the Fermi velocity, and $\sigma_i$'s are the Pauli matrices in the pseudospin basis. The electron's coupling to acoustic phonon thus is simply captured by setting $\hbar{\bm k} \to \hbar {\bm k} + e\tau {\bm A}({\bm r},t)$. Following the standard Kubo's formalism, the second order current in Dirac material can be obtained by evaluating the correlation function of three current operators $\hat {\bf j} = -(e/\hbar)\partial_{\bf k} \hat{\cal H} =-e v_F(\tau \hat\sigma_x,\hat \sigma_y)$. Accordingly, the  nonlinear susceptibility reads (see the corresponding Feynman diagram in Fig~\ref{fig2}(a))
\begin{align}\label{eqn:chiabc} 
&\chi^{(2)}_{abc}({\bm q}_1,iq_{1n},{\bm q}_2,iq_{2n}) 
= \frac{1}{2}\sum_{{\cal P}}\frac{1}{S}\sum_{\bm k} \frac{1}{\beta} \sum_{ik_n} \text{tr}\big[ 
\hat{j}_a 
\nonumber\\& \times
\hat G({\bm k},ik_n)
\hat{j}_b
\hat G({\bm k}+{\bm q}_1,ik_n+iq_{1n})
\hat{j}_c
\nonumber\\& \times
\hat G({\bm k}+{\bm q}_1+{\bm q}_2,ik_n+iq_{1n}+iq_{2n})
\big],
\end{align}
where $\hat G({\bm k},ik_n)= [ik_n-\hat {\cal H}({\bm k})]^{-1}$ is the fermionic Green's function in the Matsubara frequency domain. Note that $\sum_{\cal P}$ stands for the intrinsic permutation symmetry for the exchange $(b,{\bm q}_1,iq_{1n})\leftrightarrow (c,{\bm q}_2,iq_{2n})$, $ik_n$ is the fermionic and $iq_{in}$ is the bosonic (phononic) Matsubara energy, $S$ is the area of the system, $\beta=1/(k_B T)$ with $T$ being the electronic temperature and $k_B$ is the Boltzmann constant. The trace operation $\text{tr}[\dots]$ sums over all discrete degrees of freedom such as spin, pseudospin and valley indices.
Performing the above summation over fermion wave vector and keeping $q$ finite is a formidable task. However, we proceed in a perturbative manner by expanding the Green's function up to the linear order in ${\bm q}$ the phonon wave vector: 
\begin{align}
&\hat G({\bm k}+{\bm q},ik_n+ iq_n) = \hat G({\bm k},ik_n+ iq_n) \nonumber\\&+ {\bm q} \cdot \partial_{\bm k} \hat G({\bm k},ik_n+ iq_n) + {\cal O}(q^2). 
\end{align}
Using the relation $\hat G \hat G^{-1}=\hat I$ with $\hat I$ being the identity matrix, we have $\partial_{\bm k} \hat G  = -\hat G \partial_{\bm k}\hat G^{-1}  \hat G $. Therefore, by definition we obtain 
\begin{align}
\partial_{k_a} \hat G   = \hat G 
\left(-\frac{\hbar \hat j_a}{e}\right) \hat G~.
\end{align}
Accordingly, we arrive at the following relation for the expansion of the Green's function
\begin{align}
&\hat G({\bm k}+{\bm q},ik_n+ iq_n) = \hat G({\bm k},ik_n+ iq_n) \nonumber\\&-\frac{\hbar}{e}\sum_a q_a   \hat G({\bm k},ik_n+ iq_n) \hat j_a \hat G({\bm k},ik_n+ iq_n) + {\cal O}(q^2). 
\end{align}
Utilizing the above equation, we expand the second order response function up to leading order in ${\bm q}_i$:
\begin{align}
&\chi^{(2)}_{abc}({\bm q}_1,iq_{1n},{\bm q}_2,iq_{2n}) 
 \approx \frac{1}{2}\sum_{{\cal P}}
\bigg\{ X_{abc}(iq_{1n},iq_{2n}) 
\nonumber\\& -\frac{\hbar}{e}
\sum_d q_{1d} Y_{abdc}(iq_{1n},iq_{2n}) 
\nonumber\\&
-\frac{\hbar}{e}
\sum_d (q_{1d}+q_{2d}) Z_{abcd}(iq_{1n},iq_{2n})
\bigg\},
\end{align}
where $X_{abc}$, $Y_{abdc}$ and $Z_{abcd}$ are diagramatically depicted in Figs. \ref{fig2}(b), (c) and (d) respectively. In the following, we write the formal expressions of these correlation functions in terms of the Green's function $\hat G$ and the current operator $\hat {\bf j}$. The rank-3 tensor response function $X_{abc}$ reads 
\begin{align} \label{eqn:Xabc} 
&X_{abc}(iq_{1n},iq_{2n}) 
= \frac{1}{S}\sum_{\bm k} \frac{1}{\beta} \sum_{ik_n} \text{tr}\big[ 
\hat{j}_a \hat G({\bm k},ik_n)\hat{j}_b
\nonumber\\& \times
\hat G({\bm k},ik_n+iq_{1n})
\hat{j}_c
\hat G({\bm k},ik_n+iq_{1n}+iq_{2n})
\big].
\end{align}
The rank-4 response function $Y_{abdc}$ is given by  
\begin{align} \label{eqn:Yabdc} 
&Y_{abdc}(iq_{1n},iq_{2n}) 
= \frac{1}{S}\sum_{\bm k} \frac{1}{\beta} \sum_{ik_n} \text{tr}\big[ 
\hat{j}_a \hat G({\bm k},ik_n)\hat{j}_b
\nonumber\\& \times
\hat G({\bm k},ik_n+iq_{1n})
\hat{j}_d
\hat G({\bm k},ik_n+iq_{1n})
\hat{j}_c
\nonumber\\ & \times
\hat G({\bm k},ik_n+iq_{1n}+iq_{2n})
\big],
\end{align}
and similarly the other rank-4 response function $Z_{abcd}$ follows     
\begin{align} \label{eqn:Zabcd} 
&Z_{abcd}(iq_{1n},iq_{2n}) 
= \frac{1}{S}\sum_{\bm k} \frac{1}{\beta} \sum_{ik_n} \text{tr}\big[ 
\hat{j}_a \hat G({\bm k},ik_n) \hat{j}_b
\nonumber\\& \times
\hat G({\bm k},ik_n+iq_{1n})
\hat{j}_c
\hat G({\bm k},ik_n+iq_{1n}+iq_{2n})
\hat{j}_d 
\nonumber\\ & \times
\hat G({\bm k},ik_n+iq_{1n}+iq_{2n})
\big]. 
\end{align}

In inversion symmetric materials, any second order homogeneous tensor vanishes identically. Therefore, we have $X_{abc}=0$ in a gapless Dirac system. In centrosymmetric Dirac materials, the second order nonlinear current is finite only with the non-local driving field wave vector (${\bm q}\neq 0$). The wave vector associated with acoustic phonons is much larger than that of photons resulting in a stronger phonon-drag process compared to the photon-drag effect. In gapped Dirac materials such as gapped graphene and single-layer TMDs, the inversion symmetry is broken and one can naturally expect a non-vanishing $X_{abc}$ if it is not forbidden by rotational symmetries.   

To evaluate $Y$ and $Z$ response function we first perform the Matsubara summation on $ik_n$ and then we implement the analytical continuation $iq_{1n} \to \hbar\omega_1+i\delta$ and $iq_{2n} \to \hbar\omega_2+i\delta$ with $\delta\to0^+$.  Finally, we analytically evaluate the summation over fermion wave vector $\bm k$ in a continuum limit, $\sum_{\bm k}  \to S\int d^2{\bm k}/(2\pi)^2$. The details of derivation for $Y$ and $Z$ response functions are given in Appendix~\ref{app1}. 

Following the perturbative treatment for long wavelength sound waves, we can estimate the nonlinear response function in the leading order in phonon wave vector $q_i$ 
\begin{align}\label{eq:chi_2_gamma}
\chi^{(2)}_{abc}({\bm q}_1,\omega_1,{\bm q}_2,\omega_2)  
& =\frac{1}{2}\sum_d \Big\{
q_{1d} \gamma_{abcd}(\omega_1,\omega_2)
\nonumber\\ &
+
q_{2d} \gamma_{acbd}(\omega_2,\omega_1)\Big\}+ {\cal O} (q^2),
\end{align}
where $\gamma_{abcd}$ is a local rank-4 tensor and is defined as $\gamma_{abcd}(\omega_1,\omega_2)
=-(\hbar/e)[Y_{abdc}(\omega_1,\omega_2) +Z_{abcd}(\omega_1,\omega_2) +Z_{acbd}(\omega_2,\omega_1)]$. Note that the second order response function vanishes at the local approximation $q=0$ owing to the inversion symmetry of isotropic gapless Dirac fermionic system. In addition, the above relation satisfies the intrinsic permutation symmetry \cite{Butcher_Cotter} $\chi^{(2)}_{abc}({\bm q}_1,\omega_1,{\bm q}_2,\omega_2) =\chi^{(2)}_{acb}({\bm q}_2,\omega_2,{\bm q}_1,\omega_1)$. In the space-time domain, the second order rectification current for the sound wave implies 
$\chi^{(2)}_{abc}({\bm Q},\Omega,-{\bm Q},-\Omega)  
=
 [\chi^{(2)}_{abc}(-{\bm Q},-\Omega,{\bm Q},\Omega) ]^\ast$ which leads to the property 
 
 \begin{align}\label{eq:gammaSym}
 \gamma_{abcd}(-\Omega,\Omega) = -\gamma^\ast_{abcd}(\Omega,-\Omega).     
 \end{align}
 
 Therefore, with no need of explicit calculation, we expect the low-frequency scaling:  
${\rm Re}[\gamma_{abcd}(\Omega,-\Omega)]\sim \Omega$ and
${\rm Im}[\gamma_{abcd}(\Omega,-\Omega)]\sim 1$. Furthermore, for different spatial indices, there are sixteen components of the rank-4 tensor quantities in 2D Dirac system. Considering the isotropic symmetry of the Dirac Hamiltonian, there are just three non-vanishing independent tensor elements namely $\gamma_{xxyy}$, $\gamma_{xyyx}$, and $\gamma_{xyxy}$ and the remaining components can be expressed in terms of those, such as $\gamma_{xxxx}=\gamma_{xxyy}+\gamma_{xyxy}+\gamma_{xyyx}$ and other elements on interchanging $x\leftrightarrow y$.
After straightforward algebraic calculations and following the Green's function technique, we obtain
\begin{align}\label{eqn:LFxxxx}
    \gamma_{xxxx}(\omega_1,\omega_2) & = -\gamma_0 \varepsilon_0\left\{ \frac{1}{\hbar\omega_1 } +\frac{1}{ \hbar\omega_\Sigma} \right\} \text{sign}(\varepsilon_F)
    \nonumber\\& \times
\frac{4 \varepsilon_F^4 }{ \left((\hbar\omega_1)^2-4 \varepsilon_F^2\right) 
   \left((\hbar\omega_\Sigma)^2-4 \varepsilon_F^2\right)},
\end{align}
with $\gamma_0= N_f e^3v_F^2/(\pi \hbar\varepsilon_0)$ where $N_f=4$ stands for the valley and spin degree of freedom and $\varepsilon_0$ is a unit of energy.
The analytical expressions for other non-vanishing matrix elements of $\gamma_{abcd}$, i.e. $\gamma_{xxyy}$, $\gamma_{xyxy}$ and $\gamma_{xyyx}$, are given in the Appendix~\ref{app2}. 
Note that all frequencies contain infinitesimal imaginary part, i.e. $\omega_i\equiv \omega_i+i0^+$ and $\varepsilon_F$ is the Fermi energy. An apparent observation from this result is that the AE current changes sign and flows in opposite direction in the electron and hole doped systems. Furthermore, this expression clearly states that the AE current vanishes when the Fermi energy approaches zero. These results are in agreement with experimental measurements of AE effect in graphene \cite{bandhu_APL2013,bandhu_APL2014,bandhu_NR2016}. The derived formulas are consistent with the literature for the light induced nonlinear phenomenon \cite{wang_PRB2016,cheng_SR2017,rostami_PRB2017}. 

In the following sections, we show that the acoustoelectric and acoustogalvanic currents can be expressed in terms of $\gamma_{abcd}(\omega_1,\omega_2)$ response function that are induced by the scalar and vector potentials, respectively. 
\section{Acoustoelectric current due to scalar potentials}\label{sec:AE}
The rectified acoustoelectric current in response to the self-consistent scalar potential reads %
\begin{align}
j^{\rm AE}_a  = \chi^{(2)}_{ann}({\bm Q},\Omega,-{\bm Q},-\Omega) V^{{\rm sc}}({\bm Q},\Omega) V^{{\rm sc}}(-{\bm Q},-\Omega),
\end{align}
where $ \chi^{(2)}_{ann}$ is a current-density-density correlation function. 
The self-consistent potential is given as a summation of bare external potentials and the induced one owing to the long range Coulomb interaction, i.e. {\em the screening effect}. For instance, up to the second order perturbation, the self-consistent scalar potential reads
\begin{align}\label{eq:Vsc}
V^{{\rm sc}} ({\bf q},\omega) 
    =V({\bm q},\omega)+ v_q \delta n^{(1)}({\bm q},\omega) + v_q \delta n^{(2)}({\bm q},\omega).
\end{align}
Here $v_q=1/(2\epsilon_0 \kappa q)$ is the Fourier transform of the Coulomb interaction in 2D with $\epsilon_0$ being the vacuum permittivity and $\kappa$ as the dielectric constant of the surrounding environment~\cite{Giuliani_Vignale_book}.
Note $\delta n^{(1)}$ and $\delta n^{(2)}$ are  the linear and second order density fluctuations, respectively. Following the standard random phase approximation (RPA) screening analysis, we obtain the self-consistent density response functions \cite{mikhailov_prb_2011,rostami_PRB2017} 
\begin{align}\label{eq:dn1}
\delta n^{(1)}({\bm q},\omega) = \chi^{(1)}_{nn}({\bm q},\omega) V^{{\rm sc}}({\bm q},\omega),
\end{align}
and
\begin{align}\label{eq:dn2}
\delta n^{(2)}({\bm q},\omega) &= \sum_{{\bm q}_1,{\bm q}_2} \sum_{\omega_1,\omega_2}\chi^{(2)}_{nnn}({\bm q}_1,\omega_1,{\bm q}_2,\omega_2) V^{{\rm sc}}({\bm q}_1,\omega_1)
\nonumber\\ &\times
V^{{\rm sc}}({\bm q}_2,\omega_2) \delta({\bm q}-{\bm q}_\Sigma)\delta(\omega-\omega_\Sigma),
\end{align}
where $\chi^{(1)}_{nn}$ and $\chi^{(2)}_{nnn}$ stand for the first and second order density response functions. By plugging Eqs.~\eqref{eq:dn1} and \eqref{eq:dn2} into Eq.~\eqref{eq:Vsc}, we obtain a self-consistent relation for $V^{{\rm sc}}$: 
\begin{align} 
&V^{{\rm sc}} ({\bm q},\omega) 
    = \frac{V({\bm q},\omega)}{\epsilon({\bm q},\omega)} +  \sum_{{\bm q}_1,{\bm q}_2} \sum_{\omega_1,\omega_2}\frac{v_q \chi^{(2)}_{nnn}({\bm q}_1,\omega_1,{\bm q}_2,\omega_2)}{\epsilon({\bm q},\omega)} 
\nonumber\\ &\times    
V^{{\rm sc}}({\bm q}_1,\omega_1)
V^{{\rm sc}}({\bm q}_2,\omega_2) \delta({\bm q}-{\bm q}_\Sigma)\delta(\omega-\omega_\Sigma),
\end{align}
where ${\bm q}_\Sigma={\bm q}_1+{\bm q}_2$, and $\omega_\Sigma=\omega_1+\omega_2$. Note the dielectric function $\epsilon({\bm q},\omega)
=1-v_q \chi^{(1)}_{nn} ({\bm q},\omega)$ is given in terms of the linear density response function $\chi^{(1)}_{nn}$. In the absence of plasmon resonance at which the dielectric function vanishes, the first term in the above relation is dominant and thus we neglect nonlinear correction due to second order density response function. Therefore, the rectified sound-induced scalar potential reads
\begin{align}
V^{{\rm sc}}({\bm Q},\Omega) V^{{\rm sc}}(-{\bm Q},-\Omega) \approx 
\frac{|V({\bm Q},\Omega)|^2}{|\epsilon({\bm Q},\Omega)|^2}.
\end{align}
The linear density response function is related to the longitudinal conductivity, i.e. $\chi^{(1)}_{nn}({\bm q},\omega)=-i(q^2/\omega) \sigma({\bm q},\omega)$\, \cite{Giuliani_Vignale_book,Rostami_AOP_2021}, and in small $q$ limit the dielectric function simplifies to
\begin{align}
 \epsilon({\bm q},\omega) \approx 1+ i \frac{v_s q }{\omega } \frac{\sigma(\omega)}{\sigma_m} ~. 
\end{align}
The characteristic conductivity is given by  $\sigma_m=2\epsilon_0\kappa v_s=\sigma_0(2\kappa/\pi\alpha)(v_s/c)$ 
with a conductivity unit $\sigma_0=e^2/4\hbar$, and $\alpha\approx1/137$ the fine structure constant. 
Noting that $V({\bm Q},\Omega) = ( \Lambda_P-iQ\Lambda_D) u_L$, we find 
\begin{align}
j^{\rm AE}_a  = \chi^{(2)}_{ann}({\bm Q},\Omega,-{\bm Q},-\Omega)  
\frac{|V({\bm Q},\Omega)|^2} 
{1+(\sigma/\sigma_m)^2} .
\end{align} 
Using the gauge invariance arguments, the second order response function (current-density-density) can be written in terms of the current-current-current correlation function like  \cite{Rostami_AOP_2021}
\begin{align}
\chi_{ann} ({\bm Q},\Omega,-{\bm Q},-\Omega)  =  \sum_{bc} \frac{Q_{b} Q_c}{\Omega^2}  \chi^{(2)}_{abc}({\bm Q},\Omega,-{\bm Q},-\Omega),
\end{align}
where $\chi^{(2)}_{abc}\sim \langle \hat j_a \hat j_b \hat j_c\rangle$ is given as the correlation function of three current operators that is diagrammatically depicted in Fig. \ref{fig2}a.

Specifically,  the AE current in terms of the longitudinal and transverse basis can be decomposed like ${\bm j}^{\rm AE} = j^{\rm AE}_{L} \hat {\bm Q} + j^{\rm AE}_{T} \hat{\bm \theta}$. Accordingly, we require to know the nonlinear response function tensor elements in the longitudinal and transverse coordinates for which we utilize $\chi^{(2)}_{abc} = \hat {\bf e}_a \cdot \chi^{(2)} :\hat {\bf e}_b\hat {\bf e}_c$ with $(\hat{\bf e}_{L},\hat{\bf e}_{T})= (\hat {\bm Q},\hat {\bm \theta})$. For the scalar potentials, the associated response will be $\chi_{aLL}^{(2)}$ where the last two indices are decided by the potential. The actual response in small $q$ limit is dictated as $\chi_{aLL}^{(2)} \sim q_L\gamma_{aLLL}$. Accordingly, only the longitudinal component contributes to the AE current due to finite $\gamma_{LLLL}$ tensor element. While the transverse component $\gamma_{TLLL}$ vanishes due to the mirror symmetry constraint of 2D Dirac system.  
\begin{figure*}[t]
    \centering
    \includegraphics[width=16cm]{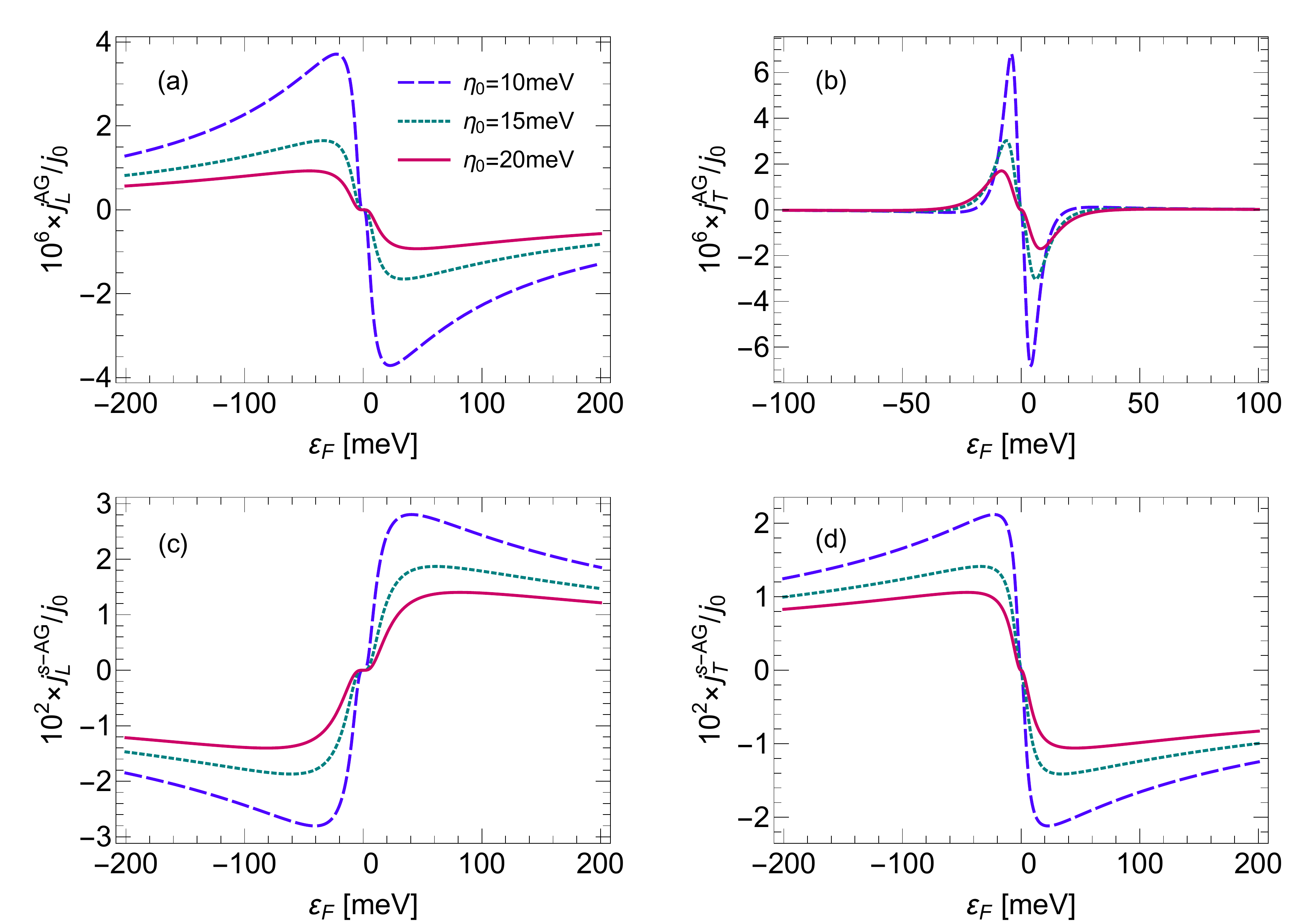}
    \caption{Fermi energy dependence of the longitudinal and transverse components of the AG current in 2D hexagonal Dirac materials such as graphene. 
    (a) and (b) indicate the longitudinal and transverse components of the uniform AG current, respectively. 
(c) and (d) illustrate  the longitudinal and transverse components of the spatially dispersive AG current, respectively. We set $\theta=0$ for the longitudinal and  $\theta=\pi/12$ for the transverse current plots. 
The plots are made for the sound frequency $\Omega=50$MHz and the scattering rate $\eta=\eta_0+0.05|\varepsilon_F|$.}
\label{fig3}
\end{figure*}
More explicitly, using Eq. (\ref{eq:chi_2_gamma}) and the symmetry constraints of $\gamma_{abcd}$ tensor in D$_6$ point group, we obtain $\chi^{(2)}_{TTT}=\chi^{(2)}_{TLL} =\chi^{(2)}_{LTL}=\chi^{(2)}_{LLT}=  0$ and %
\begin{align}\label{eq:chi_LandT}
\chi^{(2)}_{LLL}({\bm Q},\Omega,\pm{\bm Q},-\Omega) &=  Q\frac{
\gamma_{xxxx}(\Omega,-\Omega) \pm \gamma_{xxxx}(-\Omega,\Omega)
}{2},
\nonumber\\
\chi^{(2)}_{LTT}({\bm Q},\Omega,\pm{\bm Q},-\Omega) &= Q \frac{
\gamma_{xyyx}(\Omega,-\Omega) \pm \gamma_{xyyx}(-\Omega,\Omega)
}{2},
\nonumber\\
\chi^{(2)}_{TLT}({\bm Q},\Omega,\pm{\bm Q},-\Omega) &= Q \frac{
\gamma_{xyxy}(\Omega,-\Omega) \pm \gamma_{xyxy}(-\Omega,\Omega)
}{2},
\nonumber\\
\chi^{(2)}_{TTL}({\bm Q},\Omega,\pm{\bm Q},-\Omega) &=
[\chi^{(2)}_{TLT} ({\bm Q},\Omega,\pm{\bm Q},-\Omega)]^\ast.
\end{align}
\section{Longitudinal and Transverse Acoustogalvanic Current}\label{sec:AG}
For the current generated by the pseudogauge field (the AG current), the scenario of screening is completely different from the acoustoelectric current. This happens due to the strain-induced vector potential, having opposite signs in two different valleys that cannot generate a net charge current in the linear response \cite{Yudhistira_prb_2019}. Accordingly,  the self-consistent pseudogauge field is equal to the external bare one: ${\bm A}^{{\rm sc}}={\bm A}$. Therefore, the rectified current induced by the pseudogauge field does {\em not} follow the conventional screening rule $\propto 1/[1+(\sigma/\sigma_m)^2]$. In the rest of this section, we discuss the unscreened rectification current induced by the pseudogauge field, which we call ``acoustogalvanic current".

For a monochromatic Rayleigh sound wave given by Eq.~(\ref{eq:u}), we prove that the AG current reads
\begin{align}
j^{(2)}_a({\bm r}) = j^{\rm AG}_a -2j^{\rm s-AG}_a \sin(2{\bm Q}\cdot{\bm r}), 
\end{align}
where it consists of a uniform AG current component
\begin{align}
 j^{\rm AG}_a = \chi^{(2)}_{abc}({\bm Q},\Omega,-{\bm Q},-\Omega)
  A_b({\bm Q},\Omega)  A^\ast_c({\bm Q},\Omega)~, 
\end{align}
as well as a spatially dispersive one  
\begin{align}
 j^{\rm s-AG}_a = \chi^{(2)}_{abc}({\bm Q},\Omega,{\bm Q},-\Omega)
  A_b({\bm Q},\Omega)  A^\ast_c(-{\bm Q},\Omega)~.
\end{align}
Here, the spatially dispersive AG current is a direct current induced by two counterpropagating surface acoustic waves that are spatially modulated with a vanishing net current after spatial integration. However, this modulation scales with the sound wavelength $\lambda_s = 2\pi v_s/\Omega $ that is comparable with typical source-drain distance $\ell_{\rm SD}\sim 100{\rm \mu m}$. In this context, this local current density can be practically probed in experiments. Similar dispersive current has been recently measured in photogalvanic response of WTe$_2$ compounds \cite{ji_NM_2019,ji_Sc2020}. 
\begin{figure*}[t]
    \centering
    \includegraphics[width=18cm]{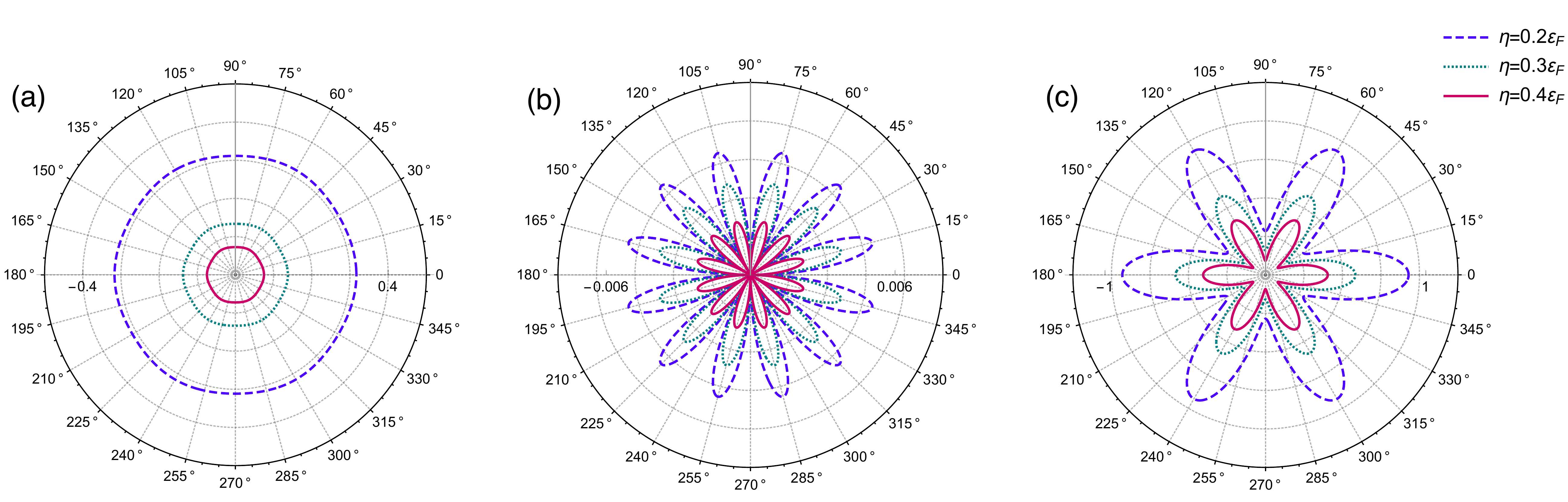}
    \caption{Angular dependence of the longitudinal AG current components in 2D hexagonal Dirac materials. (a) Longitudinal AG current $10^6\times j^{\rm AG}_L/j_0$
    (b) Transverse AG current $10^6\times j^{\rm AG}_T/j_0$ and (c) spatially dispersive longitudinal AG current $10^2\times j^{\rm AG}_L/j_0$. Note that the spatially dispersive transverse current $j^{\rm AG}_T$ holds the same anisotopic profile as in panel (b) but with stronger magnitude. As seen the uniform longitudinal current is quite isotropic while the transverse and specially dispersive current components are highly anisotropic. We set the Fermi energy $\varepsilon_F=200$meV and the sound frequency $\Omega=50$MHz.}
    \label{fig4}
\end{figure*}
By utilizing Eq. (\ref{eq:chi_2_gamma}) and Eq.~ (\ref{eq:gammaSym}), we simply find that $\chi^{(2)}_{abc}({\bm Q},\Omega,-{\bm Q},-\Omega) = \sum_d Q_d {\rm Re} [\gamma_{abcd}(\Omega,-\Omega)]$ while $\chi^{(2)}_{abc}({\bm Q},\Omega,{\bm Q},-\Omega) =i \sum_d Q_d {\rm Im} [\gamma_{abcd}(\Omega,-\Omega)]$. Afterward, using  Eq.~(\ref{eq:chi_LandT}), we write the current components in the longitudinal and transverse coordinates. Accordingly, the uniform and spatially dispersive components of the AG currents are given by 
\begin{align}\label{eq:jAG}
j^{\rm AG}_{L,T}
  = j_0 
{\rm Re}[\Pi_{ L,T}(\Omega,\varepsilon_F,\theta)],
\end{align}
%The spatial dispersive current component is proportional to the $Q(\gamma_{abcd}({\bm Q},\Omega) + \gamma_{abcd}({\bm Q},-\Omega)$. Therefore, it will be contributed by the imaginary part of $\Pi_{L,T}$ using Eqs.~\eqref{eq:chi_2_gamma} and \eqref{eq:gammaSym}. 
and
\begin{align}\label{eq:jsAG}
j^{\rm s-AG}_{ L,T}=j_0{\rm Im}[\Pi_{ L,T}(\Omega,\varepsilon_F,\theta)] 
\end{align}
where the dimensionless parameters $\Pi_{L,T}$ are given as follows 
\begin{align}
\Pi_{L}  =  \bar\gamma_{xxxx}(\Omega,-\Omega) \cos^2(3\theta)
+ \bar\gamma_{xyyx}(\Omega,-\Omega) \sin^2(3\theta),
\end{align}
and
\begin{align}
\Pi_{T} 
=[\bar\gamma_{xyyx}(\Omega,-\Omega)
-\bar\gamma_{xxxx}(\Omega,-\Omega)]\sin(6\theta),
\end{align}
where $\bar\gamma_{abcd}= \gamma_{abcd}/\gamma_0$ and $j_0= \gamma_0 A^2_0 Q I_s/4 I_0$.
Note that the sound intensity is given by $I_s =I_0 (Q u_L)^2$ with a characteristic 2D sound intensity $I_0 = \rho_{2D} v_s^3$ where $\rho_{2D}$ is the mass density of the 2D Dirac material. We should mention that the spatially dispersive (non-uniform) direct current can be also generated in response to the scalar potential, however it will be screened and fully isotropic and longitudinal. 
We highlight Eq.~(\ref{eq:jAG}) and Eq.~(\ref{eq:jsAG}) as the central results of our work for the acoustogalvanic currents in 2D hexagonal Dirac materials. 
The two striking outcomes of our study are  (i) the existence of a  transverse component of the  acoustogalvanic current, and (ii) the existence of an anisotropic dispersive component of that current. Both contributions are shown in Fig.~\ref{fig4} and discussed below. We recall that the transverse AE current is absent in the conventional 2D electron gas \cite{falko_PRB1993}.

\section{Numerical Results and Discussion }\label{sec:results}

Before presenting the quantitative results, it is worth highlighting four qualitative outcomes: ($i$) The AG current consists of both longitudinal and transverse components which depend on the direction of sound propagation $\theta$. The longitudinal AG current is the sum of two contributions  $j_1\cos^2(3\theta)$ and $j_2\sin^2(3\theta)$, while the transverse one  scales as $j_3\sin(6\theta)$. ($ii$) The $j_1$ contribution stands for the longitudinal AG current that is driven by the longitudinal pseudogauge field $A_{ L}$. The dynamical longitudinal vector potential describes a sound-induced pseudo-electric field $E\sim\Omega A_{ L}$ that results in a rectification current $j_{1}\sim E E^\ast$. ($iii$) The $j_2$ longitudinal current is driven by a transverse pseudogauge field $A_{T}$ which describes a pseudo-magnetic field $B\sim Q A_{ T}$ that results in a rectification current $j_{2}\sim B B^\ast$. ($iv$) The $j_3$ transverse current is driven by the interference of the longitudinal and transverse components of the pseudogauge field ($j_{3}\sim A_{ L} A^\ast_{ T}+c.c.$) and thus scales as $E B^\ast$.

In Fig.~\ref{fig3}, we illustrate the Fermi energy dependence of the different components of the AG current  at the sound frequency $\Omega = 50$MHz and at maximum angular variation $\theta = 0$ for the longitudinal component and $\theta = \pi/12$ for the transverse component.  Note that the uniform AG current is proportional to the real part of $\Pi$, while the spatially dispersive AG  current is proportional to the imaginary part of $\Pi$.  An immediate observation is that the AG current has opposite signs in electron and hole doped systems: this is   in agreement with experimental results \cite{bandhu_NR2016}. 
By comparing Eq.~(\ref{eq:jAG}) with Eq.~(\ref{eq:jAE}), we notice that ${\rm Re}[\Pi_{L}]\propto \mu \Gamma_s$:  therefore, the opposite sign of ${\rm Re}[\Pi_{L}]$ in electron and hole doped systems is consistent with the opposite signs of electron and hole mobility. A similar analogy works also for the spatial dispersive case for which the attenuation (dissipation) parameter is proportional to the ${\rm Im}[\Pi_{L}]/\mu$.  
There are two Lorentzian maxima in the absolute value of  ${\rm Re}[\Pi_L]$ and ${\rm Im}[\Pi_{L,T}]$ at the Fermi energy $\varepsilon_F\sim \pm \eta_0$. The case of ${\rm Re}[\Pi_T]$ is slightly different and the maxima occur at  $\varepsilon_F\sim \pm \eta_0/2$, where $\eta$ is discussed below. 

When the frequency is low, which is the case for sound waves, disorder must be taken into account and therefore a many-body analysis of the nonlinear response function $\chi^{(2)}_{abc}$ is required. However, in the highly doped regime ($|\varepsilon_F|\gg \hbar\omega_i$), 
disorder effect can be simply modeled by introducing a phenomenological relaxation rate $\eta$ via the replacement $\omega_i\to \omega_i + i\eta$.

In principle, the relaxation rate depends on the Fermi energy and the character of this dependence is different for different scattering mechanisms. At low temperature,  impurity scattering is the dominant source of scattering. At large doping and for short range impurity scattering we have $\eta\sim |\varepsilon_F|$ while for  long range scattering from charged impurities we get $\eta\sim 1/|\varepsilon_F|$ \cite{Ando_2006,Nomura_prl_2006,DasSarma_RMP_2011, Tielrooij_np_2013,rostami_cappelluti_2017,Rostami_npj2d_2021}. 
%We are thus led  to consider the following phenomenological form $\eta = \eta_0 + g |\varepsilon_F|$ where $\eta_0$ is the constant relaxation rate at $\varepsilon_F=0$ and $g$ is a dimensionless parameter characterizing disorder scattering strength. 
%
For instance, in the absence of charged impurity, we consider a more realistic phenomenological form $\eta = \eta_0 + g |\varepsilon_F|$ where $\eta_0$ is the constant relaxation rate at $\varepsilon_F=0$ and $g$ is a dimensionless parameter characterizing disorder scattering strength. 
At high doping $|\varepsilon_F|\gg\hbar\Omega$, the AG response function %with a more realistic form for $\eta$ {\bf What is this more realistic form of $\eta$??} 
shows negative slope owing to the larger scattering rate $\eta$.
%{\bf (larger scattering cross-section?)}. 
Qualitatively, the longitudinal response drops as 
${\rm Re}[\Pi_{L,T}]\sim 1/\eta^2 \sim 1/\varepsilon^2_F$
and ${\rm Im}[\Pi_{L,T}]\sim 1/\eta \sim 1/|\varepsilon_F|$. %   

The acoustogalvanic effect originates from the pseudogauge field, which inherits the three-fold symmetry of the hexagonal lattice in 2D Dirac materials like graphene and TMD families. This three-fold symmetry manifests itself in the angle dependent factors in the second order current in response to the sound-induced pseudogauge field. The transverse current is highly anisotropic and  depends on the  direction of propagation of the wave  as $j_T\sim \sin(6\theta)$ regardless of other parameters. However, the longitudinal current contains two terms $j_L\sim \sin^2(3\theta)$ and $j_L\sim \cos^2(3\theta)$ weighted by two different tensor elements of $\gamma_{abcd}$. In this regard, we depict the polar plots of longitudinal and transverse AG currents that illustrate the angular dependence of the sound propagation in Fig. \ref{fig4}. The  uniform longitudinal AG current is almost isotropic as seen in Fig. \ref{fig4}a owing to the fact that the $\sin^2(3\theta)$ and $\cos^2(3\theta)$ terms contribute almost equally. However, the transverse AG current is highly anisotropic, consistent with $\sin(6\theta)$ dependence that is evident in  Fig.~\ref{fig4}b. As seen in Fig.~\ref{fig4}c the spatially dispersive longitudinal AG current is strongly anisotropic because the $\sin^2(3\theta)$ and $\cos^2(3\theta)$ terms contribute unequally.

Finally, we compare the amplitude of the AG current to the experimental measurements of conventional AE current in graphene \cite{bandhu_NR2016,bandhu_APL2013,bandhu_APL2014,Poole2017}.  For instance, the peak current $\sim 11 $nA is measured at sound frequency $\Omega\sim 32$MHz in a graphene device with the width $\sim 3$mm leading the AE current density $j_{exp}\sim 3.6$nA/mm \cite{bandhu_NR2016}. The maximum sound intensity is reported to be $I_{s}\sim 0.3$W/m \cite{bandhu_NR2016}. We estimate the 2D mass density as $\rho_{2D}=d_{gr}\rho_{3D}$ where $\rho_{3D}=2267{\rm  kg/m^3}$ is the density of graphite and $d_{gr}\sim 1{\rm \AA}$ is the  effective thickness of the graphene layer. With sound velocity $v_s\sim 4\times 10^3$m/s, we find $I_0= \rho_{2D} v^3_s$ and therefore we obtain $I_{s}/I_0\sim 2\times 10^{-5 }$. Accordingly, for $\beta\sim 3$, $v_F\sim 10^6$m/s and lattice constant $a\sim 0.246$nm, we find $j_0  \sim 5\times 10^6$nA/mm. For $\theta=0$ (i.e., for a sound wave propagating in the zig-zag direction),  with Fermi energy $\varepsilon_F\sim 50$meV and scattering rate $\eta\sim 20$meV, we evaluate $\Pi_L$ and then we estimate the uniform longitudinal AG currents $|j^{\rm AG}_L|\sim 3~{\rm {nA}/{mm}}$. Thus, the calculated uniform AG current $j^{\rm AG}_L$ is on the order of the measured value. 

\section{Summary} 
We have discussed  the acoustogalvanic (AG) effect in 2D hexagonal Dirac materials using the Kubo formalism. Apart from the self-consistently screened deformation potential, sound propagation in Dirac materials  induces pseudo electromagnetic fields which are not subject to  screening. We have analyzed both the uniform and the spatially dispersive components of the AG current. We identify an anisotropic uniform current transverse to the sound propagation direction and a highly anisotropic profile of the spatial dispersive AG currents. The AG response changes sign in going from electron to hole-doped systems, which is consistent with expectations for the standard AE effect. While our calculations have been performed for graphene, our formalism can be easily adapted to other 2D materials such as transition metal dichalcogenides and their heterostructures. Furthermore, our results provide a direction to design future experiments to explore fundamental aspects as well as  applications of transverse and spatially dispersive acoustoelectric currents in graphene.

\section{Note}
After the publication of our manuscript, we became aware of a simultaneously experimental work by Pai Zhao et al.~\cite{Zhao_PRL2022} where authors measured acoustoelectric effects and acoustically generated Hall voltage in graphene due to the sound-induced synthetic gauge-field that resemble our theoretical predictions.

\section{Acknowledgments}
This work was supported by Nordita and the Swedish Research Council (VR 2018-04252). Nordita is supported in part by Nordforsk. We thank E. Cappelluti for carefully reading the manuscript and his useful comments. 
 
\bibliography{refs}

\begin{thebibliography}{55}
\bibitem [{Xmi()}]{Xmisc}%
  \BibitemOpen
  \href@noop {} {}\bibinfo {note} {There is an inconsistency in literature,
  e.g. Ref. [20] and Ref. [21], for the overall minus sign in Eq. (1) that can
  be due to different convention for the coordinate.}\BibitemShut {Stop}%
\end{thebibliography}

\onecolumngrid
\appendix
\section{Detailed calculation of $\gamma_{abcd}$ tensor elements using Kubo's formalism}
\label{app1}
\subsection{Calculation for $Y_{abdc}$ quantity}
According to Eq.~\eqref{eqn:Yabdc}, the fourth rank tensor quantity $Y_{abdc}$ is given by (see Feynman diagram Fig.~\ref{fig2}c) 
\begin{align} \label{eq:a1}
&Y_{abdc}(iq_{1n},iq_{2n}) 
= \frac{1}{S}\sum_{\bm k} \frac{1}{\beta} \sum_{ik_n} \text{tr}\big[ 
\hat{j}_a \hat G({\bm k},ik_n)\hat{j}_b
\hat G({\bm k},ik_n+iq_{1n})
\hat{j}_d
\hat G({\bm k},ik_n+iq_{1n})
\hat{j}_c
\hat G({\bm k},ik_n+iq_{1n}+iq_{2n})
\big],
\end{align}
where $\hat{j}$ is the current vertex, $\hat{G}$ is the Fermionic Green's function, $\beta$ refers to the inverse of the temperature. To find the solution, it is convenient to express the equation in the band basis which is represented by $\vert {\bm k}, \lambda \rangle$, where $\lambda \equiv \pm$ having $(+)$ sign for the conduction band, and $(-)$ sign for the valence band. Using the relations,
\begin{align}\label{eq:a2}
& \langle \lambda_i \vert \hat{j}_\alpha \vert \lambda_j \rangle = j_\alpha^{\lambda_i \lambda_j}, \nonumber \\
& \langle \lambda_i \vert \hat{G} \vert \lambda_j \rangle = \frac{\delta_{\lambda_i \lambda_j}}{ik_n - \varepsilon_{\bm k}^{\lambda_i}},
\end{align}
Here, the band dispersion for 2D Dirac material is $\varepsilon_{\bf k}^{\lambda} = \lambda \hbar v_F |{\bm k}|$, where $\lambda = \pm$ for the conduction and valence bands, and $v_F$ is the Fermi velocity.
Considering Eq. (\ref{eq:a2}), we rewrite the tensor quantity in Eq. (\ref{eq:a1}) as follows
\begin{align} 
Y_{abdc}(iq_{1n},iq_{2n}) &= \frac{1}{S}\sum_{\bf k} \frac{1}{\beta} \sum_{ik_n}\sum_{\lambda_i = \pm} j_a^{\lambda_1\lambda_2} j_b^{\lambda_2\lambda_3} j_d^{\lambda_3 \lambda_4} j_c^{\lambda_4\lambda_1}\frac{1}{ik_n - \varepsilon_{\bm k}^{\lambda_2}} \frac{1}{ik_n + i q_{1,n} - \varepsilon_{\bm k}^{\lambda_3}}
\nonumber\\&\times
\frac{1}{ik_n + i q_{1,n} - \varepsilon_{\bm k}^{\lambda_4}} \frac{1}{ik_n + i q_{1,n} + i q_{2,n} - \varepsilon_{\bm k}^{\lambda_1}}.
\end{align}
To stratify it further, first we perform the Matsubara summations over the Green's functions which give
\begin{align}
Y_{abdc}(\omega_{1},\omega_{2}) &= \frac{1}{S}\sum_{\bf k} \frac{1}{\beta} \sum_{\lambda_i = \pm} \frac{j_a^{\lambda_1\lambda_2} j_b^{\lambda_2\lambda_3} j_d^{\lambda_3 \lambda_4} j_c^{\lambda_4\lambda_1}}{\omega_1 + \omega_2 - \varepsilon_{\bm k}^{\lambda_1\lambda_2}} \bigg\{ \frac{1}{\omega_1 - \varepsilon_{\bm k}^{\lambda_4\lambda_2}} \bigg( \frac{n_F(\varepsilon_{\bm k}^{\lambda_2}) - n_F(\varepsilon_{\bm k}^{\lambda_3})}{\omega_1 - \varepsilon_{\bm k}^{\lambda_3 \lambda_2}} + \frac{n_F(\varepsilon_{\bm k}^{\lambda_3}) - n_F(\varepsilon_{\bm k}^{\lambda_4})}{\varepsilon_{\bm k}^{\lambda_4 \lambda_3}} \bigg) 
\nonumber \\
& +  \frac{1}{\omega_2 - \varepsilon_{\bm k}^{\lambda_1\lambda_4}} \bigg( \frac{n_F(\varepsilon_{\bm k}^{\lambda_3}) - n_F(\varepsilon_{\bm k}^{\lambda_4})}{\varepsilon_{\bm k}^{\lambda_4 \lambda_3}} + \frac{n_F(\varepsilon_{\bm k}^{\lambda_3}) - n_F(\varepsilon_{\bm k}^{\lambda_1})}{\omega_2 - \varepsilon_{\bm k}^{\lambda_1 \lambda_3}} \bigg) \bigg\},
\end{align}
having $n_F(x) = [e^{\beta(x-\mu)}+1]^{-1}$ is the Fermi-Dirac distribution function, and $\varepsilon_{\bm k}^{\lambda_i \lambda_j} = \varepsilon_{\bm k}^{\lambda_i} - \varepsilon_{\bm k}^{\lambda_j}$ is the difference between the dispersion of two bands. It is to be noted that we here use the shorthand notation $iq_{j,n} = \omega_j + i \delta = \omega_j$. Now, we will calculate the tensor quantities for different combinations of the Cartesian indices.
%
%
%\subsubsection{Calculation for $Y_{xxyy}(\omega_1,\omega_2)$}
%
First, the conductivity tensor $Y_{xxyy}(\omega_1,\omega_2)$ is given by
\begin{align}
Y_{xxyy}(\omega_{1},\omega_{2}) &= \frac{1}{S}\sum_{\bf k} \frac{1}{\beta} \sum_{\lambda_i = \pm} \frac{\mathcal{F}_{xxyy}^{1234}}{\omega_1 + \omega_2 - \varepsilon_{\bm k}^{\lambda_1\lambda_2}} \bigg\{ \frac{1}{\omega_1 - \varepsilon_{\bm k}^{\lambda_4\lambda_2}} \bigg( \frac{n_F(\varepsilon_{\bm k}^{\lambda_2}) - n_F(\varepsilon_{\bm k}^{\lambda_3})}{\omega_1 - \varepsilon_{\bm k}^{\lambda_3 \lambda_2}} + \frac{n_F(\varepsilon_{\bm k}^{\lambda_3}) - n_F(\varepsilon_{\bm k}^{\lambda_4})}{\varepsilon_{\bm k}^{\lambda_4 \lambda_3}} \bigg) 
\nonumber \\
& +  \frac{1}{\omega_2 - \varepsilon_{\bm k}^{\lambda_1\lambda_4}} \bigg( \frac{n_F(\varepsilon_{\bm k}^{\lambda_3}) - n_F(\varepsilon_{\bm k}^{\lambda_4})}{\varepsilon_{\bm k}^{\lambda_4 \lambda_3}} + \frac{n_F(\varepsilon_{\bm k}^{\lambda_3}) - n_F(\varepsilon_{\bm k}^{\lambda_1})}{\omega_2 - \varepsilon_{\bm k}^{\lambda_1 \lambda_3}} \bigg) \bigg\},
\end{align}
where $\mathcal{F}_{xxyy}^{1234} =j_x^{\lambda_1\lambda_2} j_x^{\lambda_2\lambda_3} j_y^{\lambda_3 \lambda_4} j_y^{\lambda_4\lambda_1} $ is the form factor. The latter quantity is only a function of angle and independent of the wave vector, thus on performing the angular integration we get
\begin{align}
\int_{0}^{2\pi} d\theta \mathcal{F}_{xxyy}^{1234} = & \int_{0}^{2\pi} d\theta \frac{\lambda_1 e^{-i\tau \phi({\bm k})} + \lambda_2 e^{i\tau \phi({\bm k})} }{2} \frac{\lambda_2 e^{-i\tau \phi({\bm k})} + \lambda_3 e^{i\tau \phi({\bm k})} }{2} \frac{\lambda_3 e^{-i\tau \phi({\bm k})} - \lambda_4 e^{i\tau \phi({\bm k})} }{2}  \frac{\lambda_4 e^{-i\tau \phi({\bm k})} - \lambda_1 e^{i\tau \phi({\bm k})} }{2}\nonumber \\
& =\frac{\pi}{4} (1 - \lambda_2 \lambda_4 + \lambda_1 \lambda_3).
\end{align}
It is evident from the expression for $Y_{xxyy}(\omega_{1},\omega_{2})$ that the quantity will yield strong contribution for the degenerate case $\lambda_3 = \lambda_4$. For this case, the integration over wave vector and the summation over band indices reduce the tensor quantity in the form
\begin{align}
    Y_{xxyy}(\omega_1,\omega_2) & = {\rm sgn}(\varepsilon_F)\frac{e^4v_F^2}{8\pi \hbar^2} \frac{16 \varepsilon_F^2 \big( - \omega_1 \omega_2^2 + \varepsilon_F^2 (3\omega_1 + \omega_2) \big)}{\omega_1 \omega_2 (\omega_1^2 - 4 \varepsilon_F^2)(\omega_2^2 - 4 \varepsilon_F^2)}.
\end{align}
Similarly, the other elements give
\begin{align}
Y_{xyyx}(\omega_1,\omega_2) & = \frac{e^4v_F^2}{8\pi \hbar^2} \frac{{\rm sgn}(\varepsilon_F)16\varepsilon_F^2 (\omega_2 - \omega_1)(\omega_1\omega_2 + \varepsilon_F^2)}{\omega_1 \omega_2 (\omega_1^2 - 4 \varepsilon_F^2)(\omega_2^2 - 4 \varepsilon_F^2)},
\end{align}
\begin{align}
Y_{xyxy}(\omega_1,\omega_2) & = \frac{e^4v_F^2}{8\pi \hbar^2} \frac{{\rm sgn}(\varepsilon_F)16\varepsilon_F^2 \big(\omega_1^2\omega_2 - \varepsilon_F^2(\omega_1 + 3\omega_2) \big)}{\omega_1 \omega_2 (\omega_1^2 - 4 \varepsilon_F^2)(\omega_2^2 - 4 \varepsilon_F^2)}.
\end{align}

\subsection{Calculation for $Z_{abcd}$ quantity}
The rank-4 quantity correspond to the Feynmann diagram Fig.~\ref{fig2}(d) can be written as
\begin{align}
Z_{abcd}(iq_{1n},iq_{2n}) 
&= \frac{1}{S}\sum_{\bf k} \frac{1}{\beta} \sum_{ik_n} \sum_{\lambda_i = \pm} \text{tr}\big[ 
\hat{j}_a \hat G({\bf k},ik_n) \hat{j}_b
\hat G({\bf k},ik_n+iq_{1n})
\hat{j}_c
\hat G({\bf k},ik_n+iq_{1n}+iq_{2n})
\hat {j}_d 
\nonumber\\&\times
\hat G({\bf k},ik_n+iq_{1n}+iq_{2n})
\big]. 
\end{align}
In the band basis representation, it becomes
\begin{align}
Z_{abcd}(iq_{1n},iq_{2n}) 
&= \frac{1}{S}\sum_{\bf k} \frac{1}{\beta} \sum_{ik_n} \sum_{\lambda_i = \pm} j_a^{\lambda_1\lambda_2} j_b^{\lambda_2\lambda_3} j_c^{\lambda_3 \lambda_4} j_d^{\lambda_4\lambda_1}\nonumber \\
& \times \frac{1}{ik_n - \varepsilon_{\bm k}^{\lambda_2}} \frac{1}{ik_n + i q_{1,n} - \varepsilon_{\bm k}^{\lambda_3}} \frac{1}{ik_n + i q_{1,n} + i q_{2,n} - \varepsilon_{\bm k}^{\lambda_4}} \frac{1}{ik_n + i q_{1,n} + i q_{2,n} - \varepsilon_{\bm k}^{\lambda_1}}. 
\end{align}
Further on performing the Matsubara frequency summations and then doing the analytic continuation, we obtain
\begin{align}
Z_{abcd}(\omega_{1},\omega_{2}) 
&= \frac{1}{S}\sum_{\bf k} \sum_{\lambda_i = \pm}\frac{j_a^{\lambda_1\lambda_2} j_b^{\lambda_2\lambda_3} j_c^{\lambda_3 \lambda_4} j_d^{\lambda_4\lambda_1}}{(\omega_1 - \varepsilon_{\bm k}^{\lambda_3\lambda_2})\varepsilon_{\bm k}^{\lambda_1\lambda_4}} 
\nonumber \\ 
& \times \bigg\{    \frac{n_F(\varepsilon_{\bm k}^{\lambda_4}) - n_F(\varepsilon_{\bm k}^{\lambda_2})}{\omega_1 + \omega_2 -\varepsilon_{\bm k}^{\lambda_4 \lambda_2}}  -    \frac{n_F(\varepsilon_{\bm k}^{\lambda_4}) - n_F(\varepsilon_{\bm k}^{\lambda_3})}{\omega_2 - \varepsilon_{\bm k}^{\lambda_4 \lambda_3}}  + \frac{n_F(\varepsilon_{\bm k}^{\lambda_1}) - n_F(\varepsilon_{\bm k}^{\lambda_3})}{\omega_2 - \varepsilon_{\bm k}^{\lambda_1 \lambda_3}}  -    \frac{n_F(\varepsilon_{\bm k}^{\lambda_1}) - n_F(\varepsilon_{\bm k}^{\lambda_2})}{\omega_1 + \omega_2 - \varepsilon_{\bm k}^{\lambda_1 \lambda_2}} \bigg\}.
\end{align}
For different spatial indices combinations, this expression reduces to the following forms:
\begin{align}
    Z_{xxyy}(\omega_1,\omega_2) & = {\rm sgn}(\varepsilon_F)\frac{e^4v_F^2}{8\pi \hbar^2} \frac{16 \varepsilon_F^2 \big( \omega_\Sigma \omega_2^2 - \varepsilon_F^2 (3\omega_1 + 2\omega_2) \big)}{\omega_2 \omega_\Sigma (\omega_\Sigma^2 - 4 \varepsilon_F^2)(\omega_2^2 - 4 \varepsilon_F^2)},
\end{align}
\begin{align}
    Z_{xyyx}(\omega_1,\omega_2) & =  \frac{e^4v_F^2}{8\pi \hbar^2} \frac{{\rm sgn}(\varepsilon_F)16\varepsilon_F^2 \big( \omega_2 \omega_\Sigma^2 + \varepsilon_F^2 (\omega_1 -2 \omega_2) \big)}{\omega_2 \omega_\Sigma (\omega_\Sigma^2 - 4 \varepsilon_F^2)(\omega_2^2 - 4 \varepsilon_F^2)},
\end{align}
\begin{align}
    Z_{xyxy}(\omega_1,\omega_2) & =  \frac{e^4v_F^2}{8\pi \hbar^2} \frac{{\rm sgn}(\varepsilon_F)16 \varepsilon_F^2 (\omega_1 + 2\omega_2)(\varepsilon_F^2 - \omega_2\omega_\Sigma)}{\omega_2 \omega_\Sigma (\omega_\Sigma^2 - 4 \varepsilon_F^2)(\omega_2^2 - 4 \varepsilon_F^2)}.
\end{align}

\section{Explicit expressions of $\gamma_{xxyy}$, $\gamma_{xyyx}$, and $\gamma_{xyxy}$}
\label{app2}
The different forms of four rank tensor quantity which is defined in the form $\gamma_{abcd}(\omega_1,\omega_2)
=-(\hbar/e)[Y_{abdc}(\omega_1,\omega_2) +Z_{abcd}(\omega_1,\omega_2) +Z_{acbd}(\omega_2,\omega_1)]$ are listed below:
\begin{align}\label{eqn:LFxxyy}
    \gamma_{xxyy}(\omega_1,\omega_2) & = \gamma_0
   \frac{  4 \varepsilon_0  {\rm sgn}(\varepsilon_F)\varepsilon_F^2 (2\hbar\omega_1 + \hbar\omega_2)\big(  \varepsilon_F^2 - \hbar\omega_1 \hbar\omega_\Sigma\big)}{\hbar\omega_1 \hbar\omega_\Sigma ((\hbar\omega_1)^2 - 4 \varepsilon_F^2)((\hbar\omega_\Sigma)^2 - 4 \varepsilon_F^2)},
\end{align}
\begin{align}\label{eqn:LFxyxy}
\gamma_{xyyx}(\omega_1,\omega_2) & = \gamma_0 \frac{ 2 \varepsilon_0 {\rm sgn}(\varepsilon_F)\varepsilon_F^2  \big( 8 \varepsilon_F^4\hbar\omega_2 +\varepsilon_F^2 \hbar^3(4\omega_1^3 + 8\omega_1^2\omega_2+4\omega_1\omega_2^2-2\omega_2^3) - \hbar^5\omega_1\omega_\Sigma(2\omega_1^3 + 3\omega_1^2\omega_2 - 2\omega_1\omega_2^2-\omega_2^3) \big)}{\hbar\omega_1 \hbar\omega_\Sigma ((\hbar\omega_1)^2 - 4 \varepsilon_F^2)((\hbar\omega_2)^2 - 4 \varepsilon_F^2)((\hbar\omega_\Sigma)^2 - 4 \varepsilon_F^2)},
\end{align}
and
\begin{align}\label{eqn:LFxyyx}
\gamma_{xyxy}(\omega_1,\omega_2) & = \gamma_0 \frac{  2 \varepsilon_0 {\rm sgn}(\varepsilon_F)\varepsilon_F^2 \big( 8\varepsilon_F^4\hbar (4\omega_1 + \omega_2)  - 2\varepsilon_F^2 \hbar^3
\big[10\omega_1(\omega^2_1+\omega^2_2) + 16\omega_2\omega_1^2 + \omega^3_2 \big]
+ \hbar^5\omega_1 \omega_\Sigma^2(2\omega_1^2 + \omega_2^2 + \omega_1\omega_2) }
{\hbar\omega_1 \hbar\omega_\Sigma ((\hbar\omega_1)^2 - 4 \varepsilon_F^2)((\hbar\omega_2)^2 - 4 \varepsilon_F^2)((\hbar\omega_\Sigma)^2 - 4\varepsilon_F^2)}.
\end{align}

\end{document}